\documentclass[final,5p,times,twocolumn,authoryear]{elsarticle}

\usepackage[colorlinks=true,citecolor=VividBlue,urlcolor=blue,linkcolor=black]{hyperref}
\hypersetup{colorlinks=true, urlcolor=VividBlue}

\usepackage{amssymb}
\usepackage{lipsum}
\usepackage{tabularx}
\usepackage{tablefootnote}

\journal{Astronomy $\&$ Computing}

\usepackage[dvipsnames]{xcolor}

\begin{document}

\begin{frontmatter}

\title{Blind Line Search System: BLiSS}

\author[moon]{Luis Abalo}
\affiliation[leiden]{ organization={Huygens-Kamerlingh Onnes Laboratory, Leiden University}, addressline={Postbus 9504}, city={Leiden}, postcode={2300 RA}, state={}, country={The Netherlands} }
\author[ua]{Graciela Sanjurjo-Ferrín}

%Orden alfabetico
\author[ua]{Jessica Planelles-Villalva}
\author[ua]{José Joaquín Rodes-Roca}
\author[ua]{José Miguel Torrejón}

\affiliation[ua]{
    organization={Instituto Universitario de Física Aplicada a las Ciencias y las Tecnologías, Universidad de Alicante},
    addressline={Ap. Correos 99},
    city={Alicante},
    postcode={E-03080},
    state={},
    country={Spain}
}

\begin{abstract}
The increasing sensitivity and spectral resolution of current and forthcoming X-ray observatories, including \textit{XRISM} and \textit{NewAthena}, are expected to reveal increasing numbers of weak and blended emission lines, motivating reproducible tools for their systematic identification. Existing workflows often rely on manual inspection or source-specific analysis pipelines, making homogeneous analyses of large datasets difficult. To address this need, we present BLiSS (Blind Line Search System), an open-source Python package for the fast, blind detection and characterization of emission-line candidates in one-dimensional X-ray spectra without requiring a prior physical continuum model. BLiSS is intended as an exploratory analysis tool that complements subsequent physical spectral modelling.
The package estimates an empirical baseline directly from the observed spectrum, identifies positive excesses, groups them into candidate regions, and characterizes them with Gaussian models. Candidate reliability is estimated by comparison with synthetic spectra using a Gaussian Mixture Model classifier. Finally, optional routines perform a simultaneous multi-Gaussian fit and associate detected features with compatible atomic transitions.
The methodology implemented in BLiSS has already enabled published spectroscopic studies and is presented here as a documented, modular, and publicly available software package. Its performance is demonstrated using \textit{Chandra}/HETGS and \textit{XRISM}/Resolve observations of the high-mass X-ray binary Vela X-1, one of the best-studied X-ray sources. BLiSS recovers the principal emission features reported in previous studies while providing a fast, reproducible, and instrument-independent workflow for exploratory line searches.
\end{abstract}


\end{frontmatter}

\section{Introduction}

High-resolution X-ray spectroscopy is one of the primary tools for studying astrophysical plasmas. The energies, strengths, and profiles of emission lines provide direct diagnostics of the ionization state, chemical composition, temperature, density, and dynamics of the emitting material, making them essential for investigating a wide range of environments, including X-ray binaries, active galactic nuclei, supernova remnants, and galaxy clusters \citep[e.g.][]{Kallman_2001a}.

The capabilities of X-ray spectrometers have improved steadily over the past two decades. The \textit{Chandra X-ray Observatory} \citep{Weisskopf_2000} and its High Energy Transmission Grating Spectrometer (HETGS; \citealt{Canizares_2005}), the \textit{XMM-Newton} observatory \citep{Jansen_2001} and its Reflection Grating Spectrometer (RGS; \citealt{denHerder_2001}), and, more recently, the \textit{XRISM} mission \citep{Tashiro_2025} and its Resolve X-ray microcalorimeter \citep{Kelley_2025}, provide spectra containing large numbers of resolved emission features. The forthcoming \textit{NewAthena} mission \citep{Cruise_2025} will further improve both the sensitivity and spectral resolution of X-ray spectroscopy, enabling the detection of weaker spectral features in increasingly large observational datasets. In parallel, modern spectroscopic studies increasingly rely on the consistent analysis of multiple observations, including monitoring campaigns, archival surveys, and phase-resolved spectroscopy. These developments increase the need for efficient, reproducible methods for identifying emission lines.% Para las densidades y temperaturas astrofísicas esperamos principalemnte líneas de emision, y por el momento nos centramos en eso

A common first step in the analysis of high-resolution X-ray spectra is the identification of candidate emission lines before detailed physical modeling. Candidate features are typically identified visually from the spectrum or from the residuals of a continuum fit, and are then added individually to the spectral model. While this approach is well suited to the analysis of individual observations, it becomes increasingly time-consuming when applied to the large datasets produced by modern X-ray observatories or to studies that require the consistent analysis of many spectra.  

This challenge is particularly evident in studies of X-ray binaries. Although BLiSS (Blind Line Search System) is designed as a general-purpose tool, it was originally developed with these systems in mind. In wind-fed X-ray binaries, the short orbital and spin timescales of the compact object allow different regions of the stellar wind and accretion environment to be sampled within individual observing campaigns. High-resolution, time- and phase-resolved (both orbital or sin) spectroscopy therefore provides a powerful probe of the structure and dynamics of these systems. As the compact object moves through the stellar wind, the observed spectrum changes, and the resulting variations in the emission-line spectrum reveal the geometry and dynamics of the circumstellar environment \citep[e.g.][]{Goldstein_2004a,Miskovicova_2016,Aftab_2019,Fogantini_2021,Sanjurjo-Ferrin_2021}. At the same time, these observations present a practical challenge: high-resolution spectra contain increasingly large numbers of emission lines, while time- and phase-resolved analyses divide the data into many individual spectra that must be analyzed in a consistent and reproducible manner. BLiSS has already been used in the analysis presented by \citet{2026arXiv260609579S}, where more than 1000 spectra were processed. Such a large-scale analysis would have been impractical using traditional manual spectroscopic techniques. In this particular analysis, BLiSS was integrated with the spectral package ISIS to automatically detect emission lines and perform Monte-Carlo analysis, allowing to detect transient FeK$\alpha$ emission lines, otherwise diluted in more extended spectral bins.

Several existing software packages address complementary aspects of this problem. Bayesian Blocks algorithms identify statistically significant deviations from the continuum without requiring predefined line positions \citep{Scargle_2013}. Blind Gaussian searches systematically scan unresolved Gaussian components across the spectrum to identify narrow spectral features and evaluate their significance \citep[e.g.][]{Pinto_2016,Kosec_2018}, while the cross-correlation approach introduced by \citet{Kosec_2021} combines automated line searches with Monte Carlo simulations to account for the look-elsewhere effect in large spectroscopic surveys. Other software packages focus on the analysis of spectral features once they have been identified. For example, \textit{Specutils} \citep{Specutils_2019} provides general tools for fitting and analysing spectra, while \textit{LIME} \citep{Fernandez_2024} offers a framework for spectral visualization, constrained profile fitting, and line measurements across a broad range of applications.

BLiSS was developed to formalize the exploratory stage of emission-line analysis within a single open-source workflow. It provides a fast, instrument-independent framework for the blind detection, characterization, ranking, and optional identification of emission-line candidates without requiring a prior physical continuum model. Unlike previous approaches, BLiSS combines automated candidate detection, empirical reliability estimation based on synthetic null spectra, and optional atomic line identification into a single, reproducible Python package.

As a representative science case, we apply BLiSS to the high-mass X-ray binary Vela X-1 \citep{Kretschmar_2021}. Phase-resolved spectroscopy of this system has revealed changes in the stellar wind associated with the accretion wake \citep{Amato_2021}, while long-term monitoring has shown that these phase-dependent signatures also vary from orbit to orbit, highlighting the importance of analysing individual observations in a homogeneous manner \citep{Abalo_2024}. We demonstrate BLiSS using both \textit{Chandra}/HETGS observations, for which a detailed spectroscopic analysis is available by \citet{Grinberg_2017}, and recent \textit{XRISM}/Resolve observations analysed by \citet{Diez_2025}. Owing to its rich emission-line spectrum and the availability of high-quality reference studies, Vela X-1 provides an ideal benchmark for evaluating the software. Our goal is not to derive new astrophysical results, but to demonstrate that BLiSS reliably recovers previously reported emission features within a reproducible analysis workflow.

The remainder of this manuscript is organized as follows. Section~\ref{section:blind_line_search_methodology} describes the BLiSS workflow and the underlying blind line-search methodology. Section~\ref{section:science_case_study_vela_x1} demonstrates the software using \textit{Chandra}/HETGS and \textit{XRISM}/Resolve observations of Vela X-1. Finally, Section~\ref{section:conclusions} summarizes the main features of the package and discusses future developments.

\begin{figure} %!h
\centering
\includegraphics[width=0.49\textwidth]{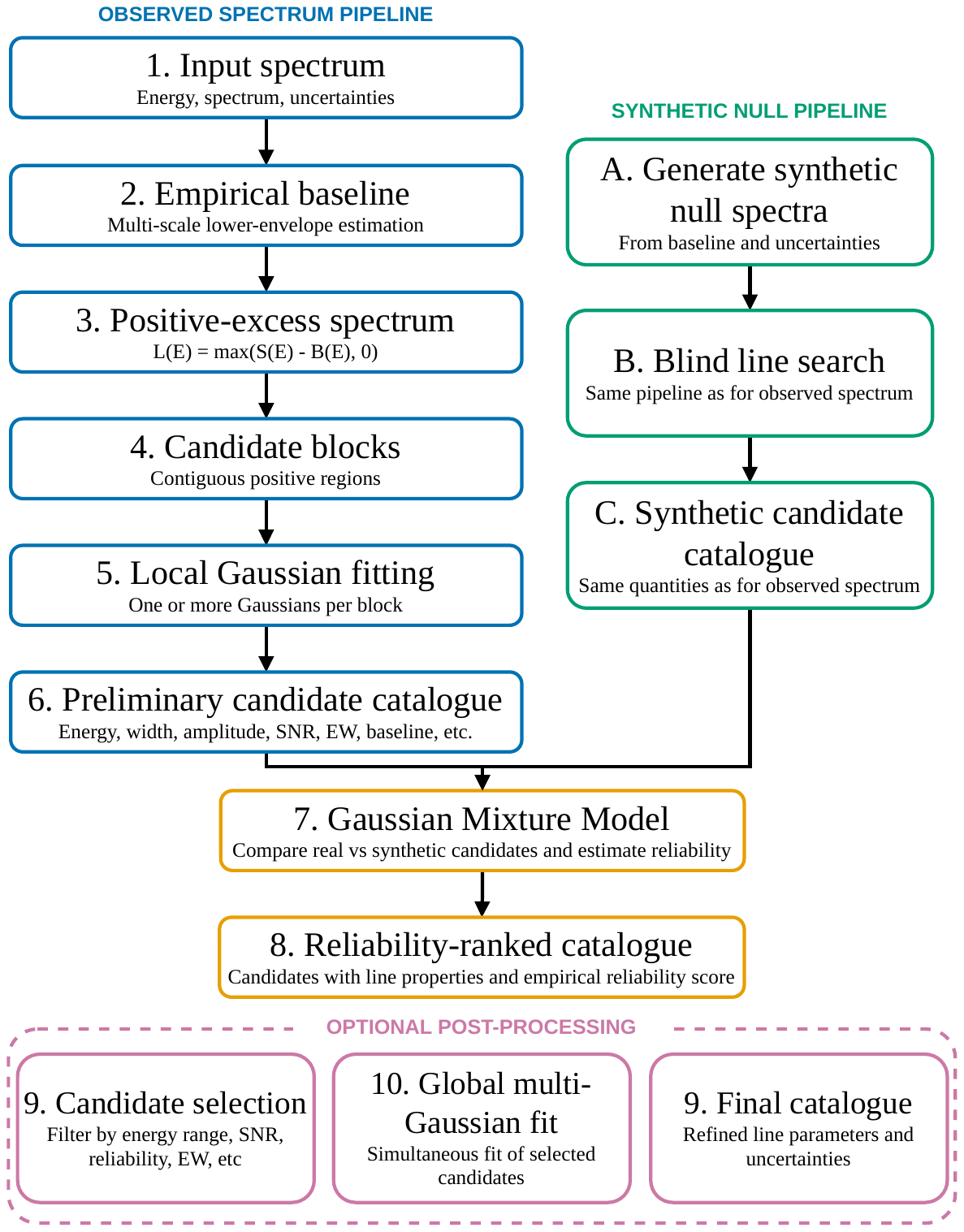}
\caption{
Schematic overview of BLiSS workflow. Blue boxes denote the processing of the observed spectrum, orange boxes the equivalent analysis performed on synthetic null spectra, green boxes the final products of the blind line-search pipeline, and gray boxes optional post-processing steps. The individual stages of the workflow are described in Sect.~\ref{section:blind_line_search_methodology}.
}
\label{figure:workflow}
\end{figure}

\section{Blind line search methodology}
\label{section:blind_line_search_methodology}

BLiSS is an open-source Python package for the automated detection, characterization, ranking, and optional identification of emission-line candidates in high-resolution X-ray spectra. Rather than relying on prior knowledge of the expected line positions or on detailed physical modelling of the continuum, BLiSS performs a blind search for localized spectral excesses and evaluates their statistical reliability through comparison with synthetic null spectra. The software is organized into independent modules corresponding to the different stages of the analysis, facilitating maintenance, future extensions, and code reuse. BLiSS is distributed through the Python Package Index (PyPI), allowing installation via \texttt{pip} (i.e. \texttt{pip install bliss-lib}), and is accompanied by online documentation and tutorial notebooks \footnote{https://github.com/xragua/BLiSS}. 

Fig.~\ref{figure:workflow} summarizes the complete BLiSS workflow. Starting from an observed X-ray spectrum and its associated uncertainties, the pipeline first constructs an empirical representation of the underlying continuum and isolates positive spectral excesses that may correspond to emission features. Candidate lines are then identified through a blind search and characterized according to their spectral and statistical properties. To assess the reliability of each candidate, the same analysis is independently applied to synthetic null spectra generated from the empirical continuum model. The distributions of candidates obtained from the observed and synthetic spectra are subsequently compared to estimate an empirical reliability score for each detection. Finally, users may apply additional selection criteria and, if desired, perform a simultaneous multi-Gaussian fit to refine the parameters of the selected lines.

The remainder of this section follows the workflow illustrated in Fig.~\ref{figure:workflow}. The data loading and preprocessing steps (stages~1--3) are described in Sect.~\ref{subsection:loading_and_preprocessing}. The blind line-search procedure (stages~4--6), together with the generation and analysis of synthetic null spectra (stages~A--C), is presented in Sect.~\ref{subsection:blind_line_search_and_synthetic_null_spectra}. The characterization, the ranking of candidate lines, and the estimation of their empirical reliability, is described in Sect.~\ref{subsection:candidate_characterization_and_ranking}. Finally, the optional identification of candidate features with known atomic transitions is discussed in Sect.~\ref{subsection:postprocessing_and_line_identification}. 

%To illustrate each stage of the workflow, we use throughout this section the three Vela X-1 spectra that are analyzed in the science case presented in Sect.~\ref{section:science_case_study_vela_x1}: two \textit{Chandra}/HETGS observations corresponding to the low- and high-hardness states analyzed by \citet{Grinberg_2017}, and one \textit{XRISM}/Resolve observation analyzed by \citet{Diez_2025}. These spectra serve solely as representative examples of the different processing stages and do not constitute the scientific validation of BLiSS, which is presented in Sect.~\ref{section:science_case_study_vela_x1}. 

\subsection{Loading and preprocessing}
\label{subsection:loading_and_preprocessing}

This subsection describes the first three stages of the BLiSS workflow (Fig.~\ref{figure:workflow}), in which the input spectrum is loaded, an empirical baseline is estimated, and the positive-excess spectrum used for the blind line search is constructed.

\paragraph{Stage 1: Input spectrum} 

BLiSS accepts one-dimensional X-ray spectra either as standard FITS spectral products or as already extracted spectra provided by the user. The latter option allows spectra generated by external analysis software, such as XSPEC \citep{Arnaud_1996} or ISIS \citep{Houck_2000}, to be analyzed without requiring the original observational products. Regardless of the input format, the data are converted into a common internal representation containing the spectral coordinate, spectral values, associated uncertainties, and, where available, the instrumental response. During this stage, BLiSS performs basic validation of the input data, discarding invalid spectral bins before proceeding with the analysis. 

\paragraph{Stage 2: Empirical baseline}

The first step of the BLiSS line-search procedure is to estimate an empirical baseline that follows the broad spectral shape. Rather than modelling the physical continuum, this baseline provides a smooth reference level from which localized emission features can be identified in a model-independent way. It is constructed from a series of sigma-clipped moving averages computed over different window sizes. At each spectral bin, BLiSS adopts the lowest of these estimates, producing a lower envelope that adapts to large-scale variations in the spectrum while remaining largely insensitive to narrow emission lines.

\begin{figure*}
\centering
\includegraphics[width=\textwidth]{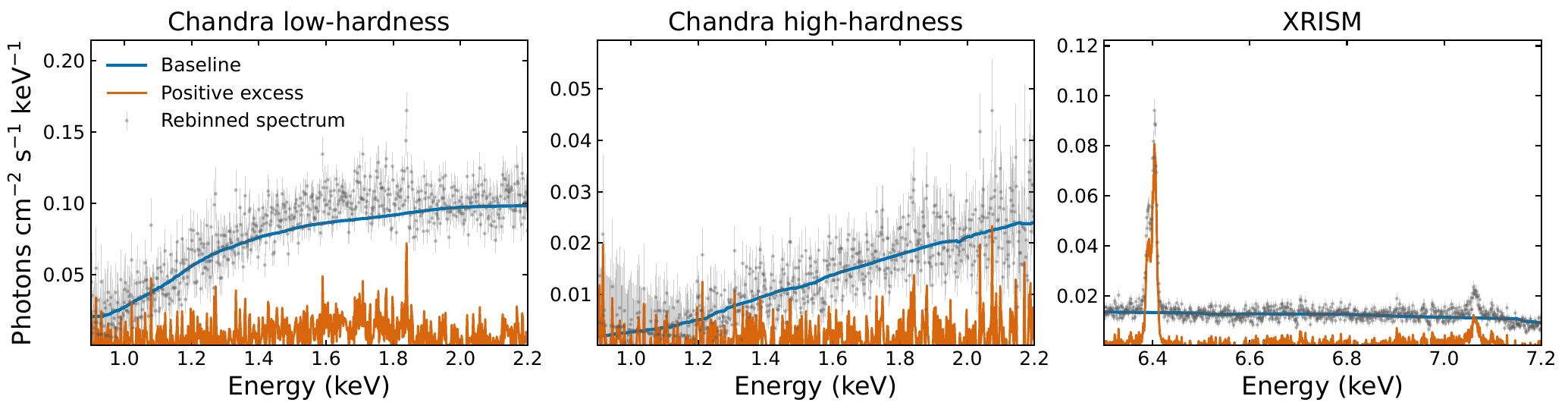}
\caption{
Empirical baseline estimation for the three Vela X-1 spectra used throughout Sect.~\ref{section:blind_line_search_methodology} to illustrate the different stages of the BLiSS workflow (Fig.~\ref{figure:workflow}). 
The left and middle panels show the low- and high-hardness \textit{Chandra}/HETGS spectra, respectively, while the right panel shows the \textit{XRISM}/Resolve spectrum.
Gray points represent the rebinned spectra with its statistical uncertainties, the blue curve shows the empirical baseline estimated by BLiSS, and the orange curve shows the positive-excess spectrum obtained after subtracting the baseline.}
\label{figure:stages1to3}
\end{figure*}

The initial lower-envelope estimate may contain narrow downward spikes that do not represent the broad spectral shape. If not corrected, these would produce artificial positive residuals in the next stage of the analysis. BLiSS therefore applies a one-sided cleaning step that selectively removes these downward excursions while preserving genuine emission features. The resulting empirical baseline serves as the reference level for the blind line search (Sect.~\ref{subsection:blind_line_search_and_synthetic_null_spectra}). 

\paragraph{Stage 3: Positive-excess spectrum}

The empirical baseline is subtracted from the observed spectrum to produce a residual spectrum (orange curve in Fig.~\ref{figure:stages1to3}). Since BLiSS is designed to search for emission features, only positive residuals are retained, while negative residuals are set to zero. Contiguous regions of positive residuals define the candidate blocks, which represent the initial search regions explored by the blind line-search algorithm in Stage~4 (Fig.~\ref{figure:workflow}).

Figure~\ref{figure:stages1to3} illustrates the preprocessing stage for the three Vela X-1 spectra. The empirical baseline is computed from 30 sigma-clipped moving-average windows spanning 3 to 50 spectral bins, followed by a one-sided downward-spike cleaning step using a 9-bin window. Despite the different continuum shapes and emission-line complexity of the three spectra, the resulting empirical baseline successfully follows the broad spectral shape while preserving the localized emission features that form the candidate blocks for the subsequent blind line search.

\subsection{Blind line search and synthetic null spectra}
\label{subsection:blind_line_search_and_synthetic_null_spectra}

This subsection describes stages~4-6 and A-C of the BLiSS workflow (Fig.~\ref{figure:workflow}). Starting from the positive-excess spectrum constructed during the preprocessing stage, BLiSS identifies localized emission features, measures their basic properties, and generates an equivalent catalog from synthetic null spectra to quantify the population of false detections expected from statistical fluctuations alone. 

\paragraph{Stage 4: Candidate blocks}

Within each positive-excess block, local maxima are identified using a peak-finding algorithm. For each local maximum, BLiSS computes its prominence, defined as the vertical contrast of the peak with respect to its surrounding local baseline, in the input spectral units. The algorithm also estimates a morphological width from the interpolated left and right positions at half prominence, and converts this width from detector bins to the physical spectral coordinate, usually keV. These peak properties are used only to initialise the subsequent Gaussian fits; the final line centroids, amplitudes, and Gaussian widths are obtained from the fitted Gaussian components.

\paragraph{Stage 5: Local Gaussian fitting}

Each candidate block is fitted with one or more Gaussian components. The detected peaks provide the initial estimates for the Gaussian centroids, while the amplitudes and widths are initialized from the local spectrum. The components are then fitted simultaneously to determine the properties of each candidate line. Performing the fit locally within each candidate block allows closely spaced emission features to be recovered without fitting the entire spectrum simultaneously.

\paragraph{Stage 6: Preliminary candidate catalogue}

The Gaussian fits from all candidate blocks are combined into a preliminary catalogue of candidate emission lines. In addition to the fitted Gaussian parameters, BLiSS records the local empirical baseline, the observed spectral value (i.e. the peak value reached by the line candidate, expressed in the input spectral units), the local noise estimate, the signal-to-noise ratio (SNR), and the equivalent width (EW) for each candidate. 

\paragraph{Stages A-C: Synthetic null spectra}

To estimate the expected population of false detections, BLiSS generates an ensemble of synthetic null spectra from the empirical baseline by adding random fluctuations consistent with the measured statistical uncertainties. These synthetic spectra preserve the spectral sampling, continuum level, and noise properties of the observation while containing no intrinsic emission features. 

Each synthetic spectrum is analyzed using exactly the same preprocessing and blind line-search procedure as the observed spectrum, producing a corresponding catalogue of candidate lines. Repeating this process for many synthetic realizations provides an empirical reference population describing the candidate lines expected from statistical fluctuations alone.

\subsection{Candidate characterization and ranking}
\label{subsection:candidate_characterization_and_ranking}

This subsection describes Stage~7 of the BLiSS workflow (Fig.~\ref{figure:workflow}), in which the preliminary catalogue is characterized and each candidate is assigned an empirical reliability score.

\paragraph{Stage 7: Candidate ranking}

The blind line search is designed to identify as many candidate emission lines as possible. As a result, the preliminary catalogue contains both genuine emission lines and candidates produced by statistical fluctuations. Since no single quantity is sufficient to distinguish between them, BLiSS combines the measured properties of each candidate into a multidimensional feature space. This information is then used to estimate an empirical reliability score for every detection. 

To estimate the reliability of each candidate, BLiSS applies an unsupervised machine-learning algorithm based on Gaussian Mixture Models (GMMs). The candidates detected in the observed and synthetic spectra are combined into the same feature space and grouped into clusters with similar properties. The number of clusters is determined automatically using the Bayesian Information Criterion (BIC). Since the synthetic catalogue represents the population of candidates expected from statistical fluctuations alone, no labelled training set is required. 

The reliability of each cluster is estimated from the relative numbers of observed and synthetic candidates that it contains. Clusters dominated by synthetic candidates are considered more likely to represent statistical fluctuations, whereas clusters containing mostly observed candidates are assigned higher reliability. This cluster reliability is then assigned to every candidate belonging to that cluster, providing a simple metric for ranking the detected emission lines.

The resulting reliability-ranked catalogue constitutes the main output of BLiSS. Depending on the scientific application, users may select candidates using quantities such as the empirical reliability score, SNR, EW, or energy range before proceeding with the optional post-processing steps.

\subsection{Post-processing and line identification}
\label{subsection:postprocessing_and_line_identification}

This subsection describes the optional final stages of the BLiSS workflow (Fig.~\ref{figure:workflow}). Starting from the reliability-ranked catalogue, users may further refine the detected candidates and, if desired, perform a tentative identification of the emission lines.

\paragraph{Stage 8: Candidate selection}

The reliability-ranked catalogue is the main output of BLiSS and can be used directly for exploratory analyses. Since the blind line search is designed to identify as many candidate lines as possible, BLiSS does not apply a fixed selection threshold. Instead, users may filter the catalogue according to the needs of their scientific application using quantities such as the empirical reliability score, SNR, EW, or energy range. 

\paragraph{Stage 9: Global multi-Gaussian fit}

The selected candidates can be refitted simultaneously over the full spectrum using an optional global multi-Gaussian fit. While the blind search fits each candidate block independently, nearby emission lines may influence one another when fitted together. The global fit therefore provides a more consistent description of the complete emission-line spectrum and improves the final line parameters. Since BLiSS uses an empirical baseline rather than a physical continuum model, quantities such as the equivalent width should be regarded as approximate. Nevertheless, fitting all selected lines simultaneously provides more consistent estimates than the preliminary local fits.

\paragraph{Stage 10: Atomic line identification}

As a final optional step, BLiSS compares the fitted line energies with an atomic transition database. Candidate lines are matched to transitions within a user-defined energy or Doppler-velocity tolerance and if included, a. 

BLiSS provides two identification modes. The first returns all compatible transitions for each detected line, allowing the user to inspect every possible association. The second assigns only the highest-ranked compatible transition to each candidate, producing a more compact catalog. Since this ranking is based only on the information contained in the \texttt{XSTAR} \citep{Kallman_2001a} atomic database, it should be regarded as a first-pass identification rather than a physical determination of the emitting ion.

\section{Science case study: Vela X-1}
\label{section:science_case_study_vela_x1}

Vela X-1 is one of the archetypal wind-fed high-mass X-ray binaries and remains a benchmark system for studying accretion from structured massive-star winds. It consists of an accreting neutron star orbiting the B-supergiant HD 77581, whose dense and variable wind produces strong absorption changes, fluorescence emission, and a rich forest of X-ray spectral lines. Despite decades of observations, the system still presents open questions related to wind clumping, photoionization, accretion wakes, transient disk formation, and the coupling between the captured flow and the neutron-star magnetosphere \citep{Kretschmar_2021}. This combination of a bright source, complex wind environment, and well-studied observational history makes Vela X-1 an ideal science case for testing BLiSS: the algorithm can be used to recover known emission-line complexes, search for additional candidate features, and assess line-detection reliability in spectra where physical complexity and instrumental sensitivity both challenge manual identification.
In this context, we apply BLiSS to three representative spectra: two hardness-resolved \textit{Chandra}/HETGS spectra and one \textit{XRISM}/Resolve spectrum. These data provide complementary tests of the algorithm, probing both the soft line-rich band accessible to gratings and the Fe K region resolved by microcalorimeter spectroscopy.

Figure~\ref{figure:stages1to3} shows the first stage of the BLiSS analysis for the three spectra considered here. The re-binned spectra are shown together with the empirical baseline estimated by BLiSS and the corresponding positive excess spectrum, defined as the part of the data lying above the baseline. In this case, the spectra were rebinned to a bin resolution of 0.002 keV in the case of Chandra spectra and 0.001 keV in the case of XRISM. To do so, we used the task \texttt{rebin\_resolution} included in the BLiSS package. Figure ~\ref{figure:stages1to3} illustrates the model-independent nature of the initial search: the empirical baseline is used only to identify candidate excesses and is not intended to represent a physical continuum model.

\subsection{Vela X-1 observed with \textit{Chandra}}
\label{sec:vela_chandra}

As a first validation case, we apply BLiSS to the hardness-resolved \textit{Chandra}/HETGS spectra of Vela X-1 extracted from ObsID 1928 and analysed by \citet{Grinberg_2017}. This observation covers orbital phases $\phi_{\rm orb}\simeq 0.21$--$0.25$, where the line of sight towards the neutron star is expected to be less affected by the large-scale accretion wake than at later orbital phases. Nevertheless, the observation shows pronounced spectral variability on kilosecond timescales, motivating the extraction of separate low- and high-hardness spectra associated with different absorption states.

This dataset provides a useful validation test for BLiSS because the published analysis combines a Bayesian-Blocks line search with detailed local modelling of individual spectral regions. The reported line inventory includes features from several ionisation stages of Si, Mg, Ne, and Fe. Here we focus on the Si, Mg, and Ne regions, and compare the emission lines detected by BLiSS with those reported by \citet{Grinberg_2017}. The published detections are therefore used as an external reference catalogue against which the probability-ranked BLiSS candidates can be assessed. The aim is not to reproduce the full physical modelling of the original work, but to test whether a blind, baseline-based search can recover the main line-like excesses and highlight state-dependent changes in the line content without imposing an a priori list of expected transitions. 

\begin{figure*}
\centering
\includegraphics[width=\textwidth]{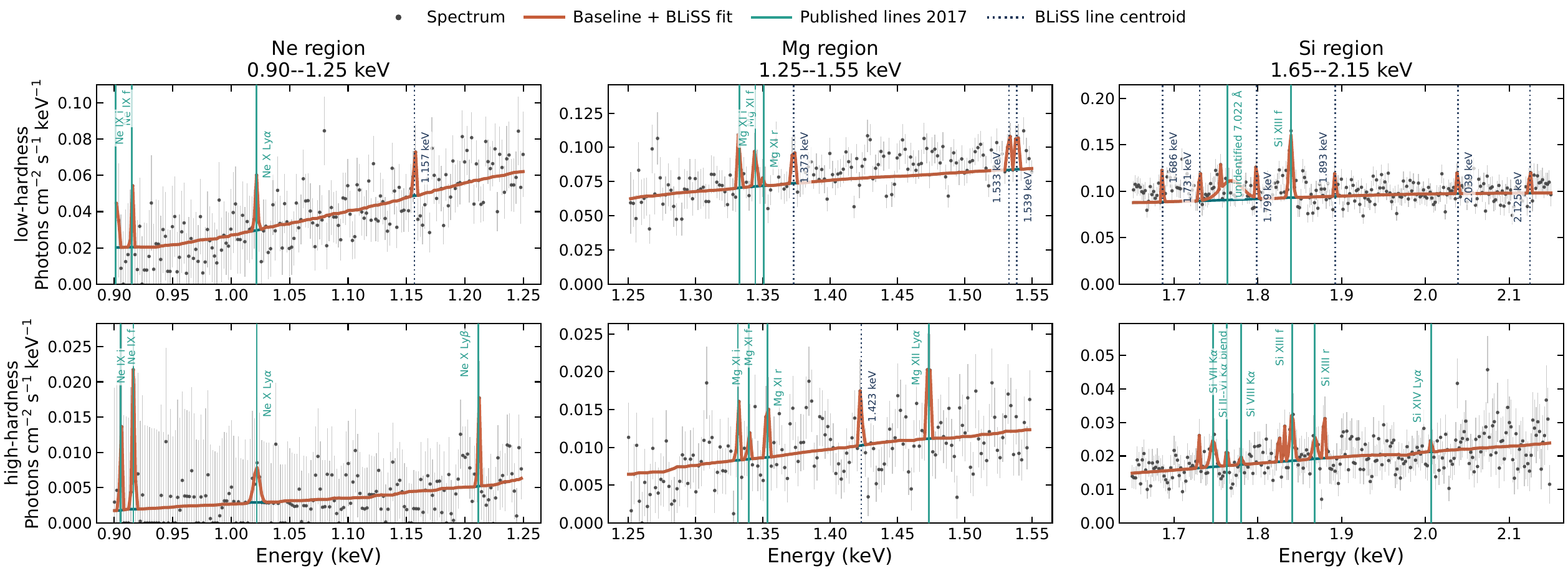}
\caption{
Comparison between the emission lines identified by BLiSS and those previously reported by \citet{Grinberg_2017} for the low- and high-hardness \textit{Chandra}/HETGS spectra of Vela X-1. Three representative spectral regions dominated by Ne, Mg, and Si emission are shown. The upper and lower rows correspond to the low- and high-hardness spectra, respectively. Grey points represent the observed spectra with its statistical uncertainties, while the orange curves show the empirical baseline together with the optional global multi-Gaussian fit obtained by BLiSS. Solid green lines indicate the line centroids reported by \citet{Grinberg_2017}, and in addition, dark blue dotted lines mark the centroids of the high-reliability emission-line candidates recovered by BLiSS.}
\label{fig:vela_chandra_victoria_global_fit_comparison}
\end{figure*}

The global fit was obtained by selecting candidate lines with \texttt{cluster\_probability} equal to 1, \texttt{relative\_power} greater than 0.1, and at least one of the signal-to-noise diagnostics, \texttt{snr\_peak}, \texttt{snr\_area}, or \texttt{snr\_amplitude}, exceeding 10. In addition, we included all lines reported by \cite{Grinberg_2017} that were detected by BLiSS as candidate features (which, at this resolution, encompassed all observed lines), even when they did not satisfy these selection criteria. A visual comparison of the BLiSS outcome is presented in Fig.~\ref{fig:vela_chandra_victoria_global_fit_comparison}, and the corresponding line parameters are given in Tables~\ref{tab:vela_x1_highhardness_lines} and \ref{tab:vela_x1_lowhardness_lines}.

BLiSS broadly reproduces the line content reported by \citet{Grinberg_2017} for the hardness-resolved \textit{Chandra} spectra of Vela X-1. The strongest agreement is found when the comparison is made by spectral region rather than line by line, since the published analysis used local, physically constrained models, whereas BLiSS decomposes the spectrum automatically into Gaussian components. BLiSS recovers the main complexes of Ne, Mg and Si, including the Mg XI triplet, Mg XII Ly$\alpha$, Si XIII f, Si XIII r, the mixture of Si II--VI K$\alpha$ and the broad unidentified feature near 7.022~\AA. The low-hardness spectrum shows systematically stronger line areas than the high-hardness spectrum, consistent with the larger line fluxes reported in the published analysis. 

The main discrepancies occur in the Ne region, where the lower number of counts and the blending of the Ne IX triplet make the individual components less robust. It must be taken into account that BLiSS does not perform an explicit automatic classification of blended line complexes. It identifies statistically significant Gaussian components and associates them with possible atomic transitions within a Doppler window, but nearby components are not grouped into a single physical blend, which should be interpreted by the user. Consequently, line-rich regions such as the He-like triplets and the low-ionization Si K$\alpha$ complex should be read with caution. This limitation highlights that BLiSS is primarily intended as a line-search tool for high-resolution spectroscopy, where future instruments will better separate neighbouring transitions.

In addition, BLiSS assigns some extra automatic identifications, especially in the Si region, which should be treated as candidate matches rather than secure physical identifications.

\subsection{Vela X-1 observed with \textit{XRISM}}
\label{sec:vela_xrism}

As a second science case, we apply BLiSS to the first published \textit{XRISM}/Resolve observation of Vela X-1, obtained as part of a simultaneous campaign with \textit{XMM-Newton} and \textit{NuSTAR} \citep{Diez_2025}. The observation was performed close to inferior conjunction of the neutron star, an orbital configuration in which the accretion and photoionisation wakes are expected to intersect the line of sight and produce strong changes in the absorbing column. The dataset shows a transition from a soft to a hard hardness-ratio interval, accompanied by an increase in the absorption column density, and therefore provides a useful test case for line searches in spectra affected by rapidly evolving wind structure.

The \textit{XRISM}/Resolve spectrum probes a complementary regime to the \textit{Chandra} grating data. While the grating spectra are especially valuable in the soft X-ray band, Resolve provides high spectral resolution in the Fe K region. This makes the observation particularly useful for testing whether BLiSS can identify closely spaced fluorescent features without prior information on their laboratory energies. The published analysis reports the Fe K$\alpha_1$/K$\alpha_2$ doublet, Fe K$\beta$, and a weak Ni K$\alpha$ component, interpreted as fluorescent emission from cold, dense clumps embedded in the ionised stellar wind.

\begin{figure*}
\centering
\includegraphics[width=\textwidth]{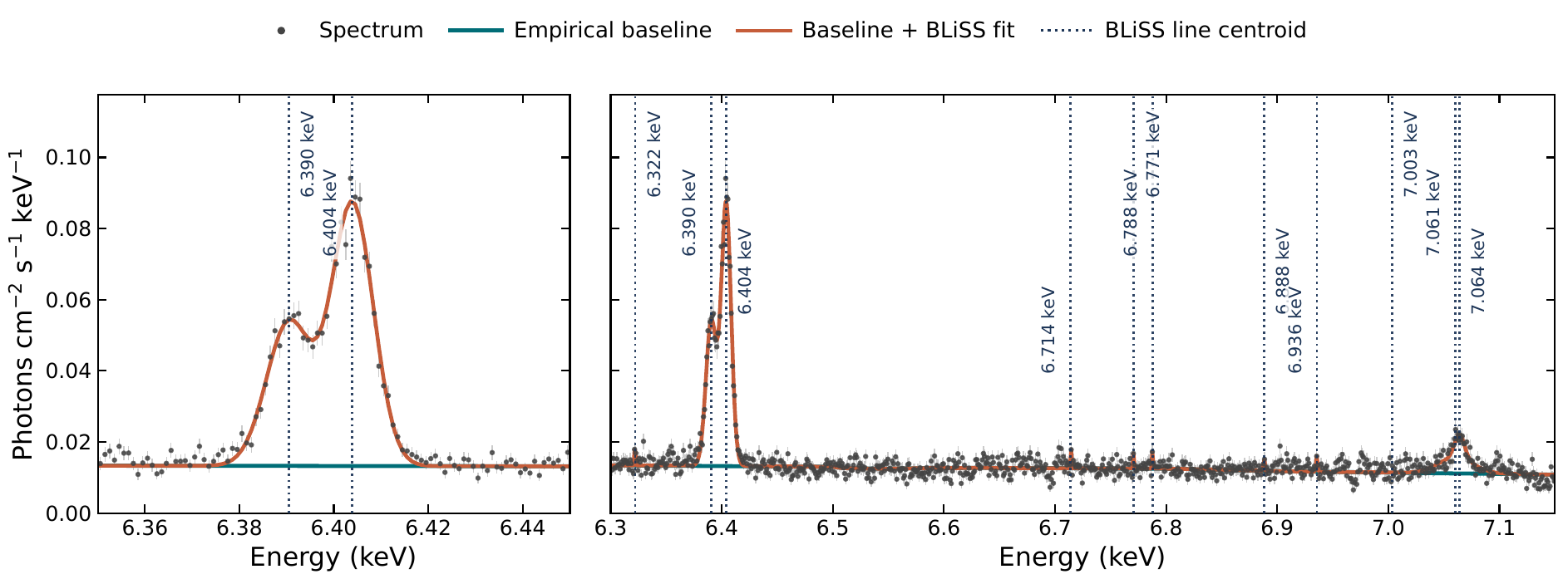}
\caption{BLiSS recovery of high-reliability emission-line candidates in the Fe K band of the \textit{XRISM}/Resolve spectrum of Vela X-1. Grey points represent the observed spectrum with its statistical uncertainties, the orange curve shows the empirical baseline together with the optional global multi-Gaussian fit, and the blue dotted lines mark the centroids of the high-reliability candidates identified by BLiSS. The left panel shows the Fe K$\alpha$ region, where BLiSS independently identifies the two components of the Fe K$\alpha_1$/K$\alpha_2$ doublet reported by \citet{Diez_2025}, without imposing either their laboratory energies or their expected separation. The right panel shows the full Fe K band, illustrating the additional high-reliability candidates recovered at higher energies around the Fe K$\beta$ and Fe~\textsc{xxvi} regions.}
\label{fig:vela_xrism_fe_zoom}
\end{figure*}

The global fit was obtained by selecting candidate lines with \texttt{cluster\_probability}=1, \texttt{relative\_power}>0.1, and at least one of the signal-to-noise diagnostics, \texttt{snr\_peak}, \texttt{snr\_area}, or \texttt{snr\_amplitude}, above 10. The \texttt{cluster\_probability} values are derived from the Gaussian Mixture Model comparison between features detected in the observed spectra and those recovered from simulated spectra.

The two strongest BLiSS components are found at $E_{\rm c}\simeq 6.39$ and $6.40$ keV, with the largest relative powers and $P_{\rm BLiSS}=1$ in both cases. Their separation is of the order expected for the Fe K$\alpha_1$/K$\alpha_2$ doublet, indicating that BLiSS does not describe the feature as a single broadened line, but resolves it into two distinct Gaussian components. As a focused validation test, Table~\ref{tab:fe_doublet_bliss_xrism_comparison} compares the BLiSS parameters of the Fe K$\alpha$ doublet with those obtained from the dedicated XRISM spectral fit of \cite{Diez_2025}. The fitted centroids and, in particular, the Gaussian widths are in close agreement, showing that BLiSS can recover closely spaced line components without imposing the expected doublet structure a priori. The absolute line areas are lower in the BLiSS fit by about 30 per cent, but the relative strength of the two doublet components is preserved.

\begin{table}
\centering
\caption{Comparison between the Fe K$\alpha$ doublet parameters recovered by BLiSS and the dedicated XRISM hard-HR spectral fit of \cite{Diez_2025}.}
\label{tab:fe_doublet_bliss_xrism_comparison}
\resizebox{\columnwidth}{!}{%
\begin{tabular}{llccc}
\hline
Line & Ref. & 
$E_{\rm c}$ (keV) & 
$\sigma$ & 
Area \\

 &  & 
 (keV) & 
$10^{-3}$ keV & 
 $10^{-4}$ ph s$^{-1}$ cm$^{-2}$\\

\hline
Fe K$\alpha_1$ 
& Diez et al. 
& $6.4041 \pm 0.0004$ 
& $4.12^{+0.38}_{-0.37}$ 
& $11.4^{+0.9}_{-1.0}$ \\

& BLiSS 
& $6.4039 \pm 0.0002$ 
& $4.45 \pm 0.16$ 
& $8.23 \pm 0.36$ \\
\hline

Fe K$\alpha_2$ 
& Diez et al. 
& $6.3905 \pm 0.0008$ 
& $4.80^{+0.78}_{-0.76}$ 
& $7.10 \pm 1.00$ \\

& BLiSS 
& $6.3904 \pm 0.0003$ 
& $4.75 \pm 0.28$ 
& $4.80 \pm 0.33$ \\
\hline
\end{tabular}%
}
\end{table}

This is precisely the regime where high-resolution spectroscopy provides a clear advantage: line complexes that appear blended at lower resolution can be decomposed into individual candidates by the blind search itself. In addition to the neutral or mildly ionised Fe K$\alpha$ doublet, BLiSS also flags weaker candidate features around the Fe XXV/Fe XXVI region and near the Fe K$\beta$ band. These detections illustrate the sensitivity of the method to weaker residual structures (see Fig. \ref{fig:vela_xrism_fe_zoom} and Table \ref{tab:vela_x1_xrism_global_fit_lines}).

\subsection{Indicative runtime}
The line-search step had modest computational cost for the spectral intervals analysed in this work. On the local workstation used for this analysis, equipped with an 8-core processor and 16 GB of RAM and running Python 3.12.2, the BLiSS line-search step required wall-clock times of 8.7 s for the Chandra high-hardness spectrum, 22.3 s for the Chandra low-hardness spectrum, and 14.0 s for the XRISM/Resolve spectrum. These timings refer only to the selected energy intervals considered here, namely the Ne, Mg, and Si regions for the Chandra spectra and the Fe region for the XRISM spectrum, and should therefore be interpreted as indicative runtime measurements rather than as a full-spectrum benchmark.

%Overall, the XRISM application provides a complementary validation to the Chandra test. While the Chandra spectra demonstrate that BLiSS can recover the main line-rich regions in lower-count, partially blended data, the XRISM spectrum shows that, when the instrumental resolution is sufficient, BLiSS can separate closely spaced components in a blind search. This supports the use of BLiSS as a discovery and pre-selection tool for the high-resolution microcalorimeter era. ##QUZÁS ESTO en conclusiones....

%%%%%%%%%%%%%%%%%%%%%%%%%%%%%%%%%%%%%%%%%%%%%%%%%%%%%%%%%%%%%%%%%%%%%%%%%%%%%%%%%%%%%%%%%%
%%%%%%%%%%%%%%%%%%%%%%%%%%%%%%%%%%%%%%%%%%%%%%%%%%%%%%%%%%%%%%%%%%%%%%%%%%%%%%%%%%%%%%%%%%
%%%%%%%%%%%%%%%%%%%%%%%%%%%%%%%%%%%%%%%%%%%%%%%%%%%%%%%%%%%%%%%%%%%%%%%%%%%%%%%%%%%%%%%%%%

\section{Summary and conclusions}
\label{section:conclusions}

We have presented BLiSS, an open-source Python package designed for the blind, reproducible detection and characterization of emission-line candidates in one-dimensional X-ray spectra. The package is intended as an exploratory analysis tool, providing a fast first-pass catalogue of candidate features before detailed physical spectral modelling. BLiSS estimates an empirical baseline directly from the observed spectrum, isolates positive excesses, fits candidate regions with Gaussian components, and ranks the resulting detections through comparison with synthetic null spectra using a Gaussian Mixture Model classifier. Optional post-processing routines allow the user to apply custom selection criteria, perform a simultaneous multi-Gaussian fit, and associate candidate features with compatible atomic transitions.

We demonstrated the performance of BLiSS using high-resolution observations of the wind-fed high-mass X-ray binary Vela X-1. In the Chandra/HETGS hardness-resolved spectra, BLiSS broadly recovers the main line-rich regions reported by \citep{Grinberg_2017}, including the Ne, Mg, and Si complexes. The low-hardness spectrum shows systematically stronger line areas than the high-hardness spectrum, consistent with the published analysis.

The XRISM/Resolve application illustrates one of the clearest strengths of BLiSS. In this case, BLiSS blindly separates the Fe K$\alpha_1$/K$\alpha_2$ doublet into two high-probability candidates, without imposing the expected laboratory energies or doublet separation. The method also flags weaker candidate structures around the Fe K$\beta$ and highly ionized Fe regions. This demonstrates that, when the instrumental resolution is sufficient, BLiSS can recover closely spaced line complexes in a blind search.

The results also illustrate the limitations of the method. BLiSS does not replace a physical continuum or plasma model, and its output should not be interpreted as a final line identification by itself. The candidate catalogue depends on the adopted binning, the empirical baseline, the local noise properties, and the user-defined selection thresholds. In blended or low-count regions, especially for He-like triplets or complex low-ionization line blends, the automatic Gaussian components should be treated as statistically motivated candidates that require subsequent physical interpretation.

On the other hand, it is worth noting that BLiSS has been deliberately developed as a scriptable, table-driven tool. Instead of aiming to replace existing spectral-analysis environments, it acts as a lightweight layer that can be integrated into more extensive workflows. The catalogue of candidate lines returned can also be exported as a simple tabular file, enabling BLiSS to be employed in loops over observations, orbital phases, hardness states, or specific energy ranges. This design choice makes it easy to interface the package with external analysis environments or other user-defined pipelines.

Overall, BLiSS provides a practical and reproducible workflow for exploratory emission-line searches in high-resolution X-ray spectra. Its main value lies in rapidly identifying, ranking, and preselecting candidate features in a homogeneous way, particularly for large datasets, phase-resolved studies, and forthcoming microcalorimeter observations. The Vela X-1 applications show that BLiSS can recover previously reported emission-line structures while providing an instrument-independent framework suited to the increasing complexity of modern X-ray spectroscopic datasets.

\section*{Acknowledgements}
The authors gratefully acknowledge Victoria Grinberg and Camille Diez for generously sharing their Vela X-1 spectra, which were essential for refining the method and illustrating the science cases presented in this work. GSF, JPV, JJRR, and JMT acknowledge the financial support from the MICIU/AEI/10.13039/501100011033 with funding from the European Union (FEDER). Project (NewAthena24-UA), reference PID2024-155779OB-C33.

\newpage

\appendix
\section{BLiSS filtered candidate-line catalogues}
\label{app:bliss_line_catalogues}

The following tables list the candidate-line parameters obtained from the optional multi-Gaussian BLiSS fits. The tentative identifications are based on the automatic matching of fitted centroids to compatible transitions within the adopted matching window. They are intended as first-pass associations and should not be interpreted as unique physical line identifications. Rows marked with $\sim$ denote parameters for which the formal uncertainties were not robust or not informative. For isolated features, local candidate fits can yield smaller formal uncertainties than the global multi-Gaussian fit, because only a limited number of parameters are allowed to vary within each candidate block. In blended regions, however, the uncertainties from both local and global fits should be interpreted with caution because neighbouring components may be correlated.

\begin{table*}
\centering
\caption{BLiSS candidate-line parameters for the Chandra high-hardness Vela X-1 spectrum. }
\label{tab:vela_x1_highhardness_lines}
\scriptsize
\setlength{\tabcolsep}{4pt}
\begin{tabular}{ccccccc}
\hline
$E_{\rm c}$ (keV) &
$\sigma$ ($10^{-3}$ keV) &
Area ($10^{-4}$ photons cm$^{-2}$ s$^{-1}$) &
S/N$_{\rm peak}$ &
Rel. power &
$P_{\rm BLiSS}$ &
Tentative ID \\
\hline

\multicolumn{7}{l}{\textbf{High hardness, Ne region (0.90--1.25 keV)}} \\
\hline
$\sim 0.91$ & $\sim 0.47$ & $\sim 0.70$ & 3.74 & 0.77 & 0.00 & Ne~IX f \\
$0.92 \pm 0.01$ & $1.04 \pm 1.03$ & $\sim 0.52$ & 1.35 & 0.83 & 1.00 & Ne~IX r \\
$\sim 1.02$ & $\sim 2.46$ & $\sim 0.31$ & 0.55 & 0.49 & 0.00 & Ne~X Ly$\alpha$ \\
$\sim 1.21$ & $\sim 0.47$ & $\sim 0.58$ & 11.57 & 0.54 & 1.00 & Ne~X Ly$\beta$ \\

\hline
\multicolumn{7}{l}{\textbf{High hardness, Mg region (1.25--1.55 keV)}} \\
\hline
$\sim 1.33$ & $\sim 0.49$ & $\sim 0.34$ & 7.89 & 0.32 & 1.00 & Mg~XI f \\
$\sim 1.34$ & $\sim 0.49$ & $\sim 0.20$ & 4.67 & 0.17 & 1.00 & Mg~XI i \\
$\sim 1.35$ & $\sim 0.48$ & $\sim 0.63$ & 14.53 & 0.27 & 1.00 & Mg~XI r \\
$\sim 1.42$ & $\sim 0.47$ & $\sim 0.48$ & 11.58 & 0.26 & 1.00 & Mg~XI \\
$\sim 1.47$ & $\sim 0.50$ & $\sim 0.84$ & 17.14 & 0.29 & 1.00 & Mg~XII Ly$\alpha$ \\

\hline
\multicolumn{7}{l}{\textbf{High hardness, Si region (1.65--2.15 keV)}} \\
\hline
$\sim 1.73$ & $\sim 0.48$ & $\sim 0.56$ & 11.93 & 0.23 & 1.00 & Si~IV \\
$1.75 \pm 0.01$ & $ 3.2 \pm 1.4$ & $0.6 \pm 0.4$ & 1.85 & 0.16 & 1.00 & Si~II--VI K$\alpha$ blend \\
$\sim 1.76$ & $\sim 0.48$ & $\sim 0.38$ & 8.36 & 0.10 & 1.00 & Si~VII K$\alpha$ \\
$\sim 1.78$ & $\sim 1.68$ & $\sim 0.10$ & 0.61 & 0.06 & 0.00 & Si~VIII K$\alpha$ \\
$\sim 1.83$ & $\sim 0.49$ & $\sim 0.59$ & 11.67 & 0.16 & 1.00 & Si~XI \\
$\sim 1.83$ & $\sim 0.47$ & $\sim 0.54$ & 11.39 & 0.22 & 1.00 & Si~XI \\
$1.84 \pm 0.01$ & $ 2.3 \pm 0.8$ & $0.8 \pm 0.4$ & 2.70 & 0.24 & 1.00 & Si~XIII f \\
$1.87 \pm 0.01$ & $2.7 \pm 2.0$ & $\sim 0.44$ & 1.22 & 0.13 & 0.00 & Si~XIII r \\
$\sim 1.88$ & $\sim 0.54$ & $\sim 0.80$ & 11.42 & 0.23 & 1.00 & Mg~XII \\
$\sim 2.01$ & $\sim 1.08$ & $\sim 0.10$ & 0.75 & 0.07 & 0.00 & Si~XIV Ly$\alpha$ \\
\hline
\end{tabular}

\vspace{1mm}
\begin{flushleft}
\footnotesize
\textit{Notes.} The tentative IDs are automatic BLiSS database matches and should be treated as candidate associations only. In blended regions, especially the Ne triplet and low-ionization Si K$\alpha$ complex, individual Gaussian components may not correspond one-to-one to physical transitions.
\end{flushleft}
\end{table*}

\begin{table*}
\centering
\caption{BLiSS candidate-line parameters for the Chandra low-hardness Vela X-1 spectrum.}
\label{tab:vela_x1_lowhardness_lines}
\scriptsize
\setlength{\tabcolsep}{4pt}
\begin{tabular}{ccccccc}
\hline
$E_{\rm c}$ (keV) &
$\sigma$ ($10^{-3}$ keV) &
Area ($10^{-4}$ photons cm$^{-2}$ s$^{-1}$) &
S/N$_{\rm peak}$ &
Rel. power &
$P_{\rm BLiSS}$ &
Tentative ID \\
\hline

\multicolumn{7}{l}{\textbf{Low hardness, Ne region (0.90--1.25 keV)}} \\
\hline
$\sim 0.90$ & $\sim 0.81$ & $\sim 0.84$ & 1.31 & 0.38 & 0.45 & Ne~IX f \\
$\sim 0.91$ & $\sim 0.45$ & $\sim 1.69$ & 5.30 & 0.45 & 1.00 & Ne~IX r \\
$1.02 \pm 0.01$ & $1.1 \pm 0.7$ & $ 0.8 \pm 0.8 $ & 1.71 & 0.34 & 0.32 & Ne~X Ly$\alpha$ \\
$\sim 1.16$ & $\sim 0.49$ & $\sim 1.61$ & 9.92 & 0.20 & 1.00 & Fe~XIX \\

\hline
\multicolumn{7}{l}{\textbf{Low hardness, Mg region (1.25--1.55 keV)}} \\
\hline
$\sim 1.33$ & $\sim 0.48$ & $\sim 2.46$ & 17.06 & 0.22 & 1.00 & Mg~XI f \\
$\sim 1.34$ & $\sim 1.04$ & $\sim 0.81$ & 2.48 & 0.15 & 0.32 & Mg~XI i \\
$\sim 1.35$ & $\sim 0.51$ & $\sim 0.19$ & 1.14 & 0.04 & 0.22 & Mg~XI r \\
$\sim 1.37$ & $\sim 0.48$ & $\sim 2.22$ & 15.17 & 0.13 & 1.00 & Mg~IX \\
$\sim 1.53$ & $\sim 0.48$ & $\sim 2.15$ & 19.23 & 0.13 & 1.00 & Fe~XXII \\
$\sim 1.54$ & $\sim 0.51$ & $\sim 1.98$ & 16.33 & 0.12 & 1.00 & Fe~XXII \\

\hline
\multicolumn{7}{l}{\textbf{Low hardness, Si region (1.65--2.15 keV)}} \\
\hline
$\sim 1.69$ & $\sim 0.73$ & $\sim 0.87$ & 4.73 & 0.16 & 1.00 & Si~VIII \\
$\sim 1.73$ & $\sim 0.53$ & $\sim 1.80$ & 11.20 & 0.14 & 1.00 & Si~V \\
$\sim 1.76$ & $\sim 0.42$ & $\sim 1.11$ & 9.40 & 0.18 & 1.00 & Si~VII \\
$1.77 \pm 0.01$ & $13 \pm 3$ & $6.7 \pm 2.1$ & 1.85 & 0.10 & 1.00 & unidentified 7.022~\AA \\
$\sim 1.80$ & $\sim 0.49$ & $\sim 2.79$ & 24.74 & 0.16 & 1.00 & Si~X \\
$1.84 \pm 0.01$ & $2.1 \pm 0.3$ & $3.6 \pm 0.8$ & 6.86 & 0.28 & 1.00 & Si~XIII f \\
$\sim 1.89$ & $\sim 0.55$ & $\sim 1.06$ & 7.74 & 0.11 & 1.00 & Si~XI \\
$\sim 2.04$ & $\sim 0.47$ & $\sim 1.68$ & 14.68 & 0.10 & 1.00 & Ni~XXVII \\
$\sim 2.12$ & $\sim 0.48$ & $\sim 1.97$ & 13.90 & 0.10 & 1.00 & Ni~XXVII \\
\hline
\end{tabular}

\vspace{1mm}
\begin{flushleft}
\footnotesize
\textit{Notes.} The tentative IDs are automatic BLiSS database matches and should be treated as candidate associations only. The low-hardness spectrum contains stronger and more blended line emission than the high-hardness spectrum, so several fitted Gaussian components may represent parts of line complexes rather than isolated physical transitions.
\end{flushleft}
\end{table*}

\begin{table*}
\centering
\caption{LiSS candidate-line parameters for the XRISM/Resolve Vela X-1 spectrum. }
\label{tab:vela_x1_xrism_global_fit_lines}
\scriptsize
\setlength{\tabcolsep}{4pt}
\begin{tabular}{ccccccc}
\hline
$E_{\rm c}$ (keV) &
$\sigma$ ($10^{-3}$ keV) &
Area ($10^{-4}$ photons cm$^{-2}$ s$^{-1}$) &
S/N$_{\rm peak}$ &
Rel. power &
$P_{\rm BLiSS}$ &
Tentative ID \\
\hline

\multicolumn{7}{l}{\textbf{XRISM/Resolve, Fe region (6.0--7.5 keV)}} \\
\hline
$\sim 6.27$ & $\sim 0.23$ & $\sim 0.11$ & 10.59 & 0.12 & 1.00 & Fe~XX \\
$\sim 6.32$ & $\sim 0.26$ & $\sim 0.17$ & 13.99 & 0.14 & 1.00 & Fe~XXI \\
$6.390 \pm 0.001$ & $4.8 \pm 0.3$ & $4.8 \pm 0.3$ & 13.60 & 0.61 & 1.00 & Fe K$\alpha_2$ \\
$6.404 \pm 0.001$ & $4.45 \pm 0.16$ & $ 8.2 \pm 0.4$ & 24.85 & 0.75 & 1.00 & Fe K$\alpha_1$ \\
$\sim 6.71$ & $\sim 0.23$ & $\sim 0.14$ & 13.56 & 0.17 & 1.00 & Fe~XXV \\
$\sim 6.77$ & $\sim 0.23$ & $\sim 0.13$ & 12.36 & 0.14 & 1.00 & Fe~XXV \\
$\sim 6.79$ & $\sim 0.29$ & $\sim 0.09$ & 7.04 & 0.16 & 1.00 & Fe~XXV \\
$\sim 6.89$ & $\sim 0.28$ & $\sim 0.09$ & 7.20 & 0.12 & 1.00 & Fe~XXV \\
$6.936 \pm 0.001$ & $0.6 \pm 0.3$ & $0.07 \pm 0.05$ & 2.54 & 0.16 & 1.00 & Fe~XXVI \\
$7.003 \pm 0.001$ & $0.01 \pm 0.01$ & $0.02 \pm 0.01$ & 46.80 & 0.23 & 1.00 & Fe~XII \\
$7.061 \pm 0.001$ & $16 \pm 3$ & $1.7^{+0.6}_{-0.5}$ & 2.20 & 0.29 & 1.00 & Fe K$\beta$ blend \\
$7.064 \pm 0.001$ & $4.5 \pm 1.1$ & $0.75 \pm 0.23$ & 3.35 & 0.33 & 1.00 & Fe K$\beta$ \\
\hline
\end{tabular}
\vspace{1mm}
\begin{flushleft}
\footnotesize
\textit{Notes.} The two strongest components near 6.4 keV correspond to the BLiSS decomposition of the Fe K$\alpha_1$/K$\alpha_2$ doublet. The weaker candidates at higher energies should be interpreted with caution: they indicate statistically selected residual structures and possible database matches, not secure physical identifications.
\end{flushleft}
\end{table*}

\newpage
\bibliographystyle{elsarticle-harv} 
\bibliography{biblio} 



%% For example:

%% \bibitem[Aladro et al.(2015)]{Aladro15} Aladro, R., Martín, S., Riquelme, D., et al. 2015, \aas, 579, A101

%%\end{thebibliography}

\end{document}